\documentclass[12pt]{article}%
\usepackage{amsmath}
\usepackage{graphicx}%
\usepackage{amsfonts}%
\usepackage{amssymb}

\begin{document}

\author{C. S. Unnikrishnan\\\textit{Gravitation Group, Tata Institute of Fundamental Research, }\\\textit{Homi Bhabha Road, Mumbai - 400 005, India}}
\title{On the Claim of the Observation of 'Macro-quantum effects' of Magnetic Vector Potential in the Classical Domain}
\date{}
\maketitle

\begin{abstract}
I present conclusive arguments to show that a recent claim of observation of
quantum-like effects of the magnetic vector potential in the classical
macrodomain is spurious. The `one dimensional interference patterns' referred
to in the paper by R. K. Varma \textit{et al} (Phys. Lett. A 303 (2002)
114--120) are not due to any quantum-like wave phenomena. The data reported in
the paper are not consistent with the interpretation of interference, or with
the topology of the Aharonov-Bohm effect. The assertion that they are evidence
of A-B like effect in the classical macrodomain is based on inadequate
appreciation of basic physical facts regarding classical motion of electrons
in magnetic fields, interference phenomena, and the A-B effect.

\end{abstract}

\bigskip\ 

\noindent E-mail address: unni@tifr.res.in\medskip



\bigskip\pagebreak 

Recently, R. K. Varma \textit{et al} reported on evidence for the observation
of the effect of magnetic vector potential in the classical domain
\cite{varmaab}. This was interpreted by the authors as equivalent to the
quantum Aharonov-Bohm (A-B) effect -- phase changes in the quantum wave
function and resulting shift in an interference pattern due to the difference
in the vector potential sampled by two possible quantum paths of the system.
The experiment by Varma \textit{et al} consisted of the observation of the
current of an electron beam from an electron gun as a function of the magnetic
flux in a solenoidal coil through which the electron beam passes without
physical contact. The leakage fields in a such a coil were apparently small
enough not to affect the beam due through \ Lorentz force; yet oscillatory
patterns were seen in the current reminiscent of the movement of an
interference fringe pattern as a function of the phase change.

The aim of this brief paper is to point out that there are serious flaws in
their interpretation, and that the observed effects could not be due to
interference effects. In fact, it had already been pointed out earlier that
the one dimensional interference pattern that forms the basis of the whole
interpretation is a result of simple classical focussing of the electron beam
in the axial magnetic field \cite{mir,unni1}. Thus there is no quantum-like
phenomena involved, and the pattern itself could be explained by well known
classical effects. The ``macro-quantum dynamical'' effects described in a
recent review article by Varma \cite{var-physrep} are due to classical
dynamics of electrons, especially secondary electrons, in static magnetic and
electric fields, but wrongly interpreted as new macroscopic quantum effects.
Since any classical probability function can be described formally as the
square of a `wavefunction', it is possible to describe these classical effects
in terms of equation involving the wavefunction. But, that is just a
quantum-like theory of classical phenomena, and does not represent a new
manifestation of quantum dynamics in the macrodomain.%

\begin{figure}
[ptb]
\begin{center}
\includegraphics[
height=2.2295in,
width=5.0237in
]%
{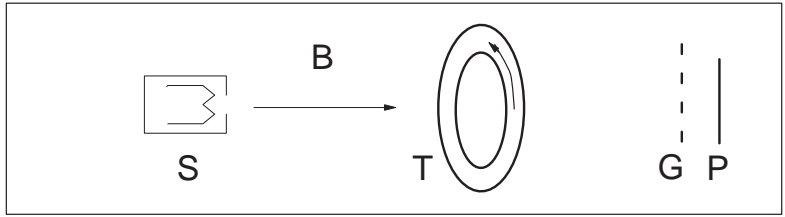}%
\caption{Schematic diagram of the experiment. S) electron source, T) toroidal
magnet, G) grid, and P) plate detector. There is a uniform axial magnetic
field B generated by coils external to the vacuum chamber. The toridal coil
confines a magnetic field. }%
\end{center}
\end{figure}

In the experiment \cite{varmaab}, Varma \textit{et al} monitored the current
of electrons from an electron gun reaching a metal plate detector placed
behind a wire grid inside \ a vacuum chamber, with a source to detector
separation of about 30 cm (see Fig. 1 for a schematic diagram). The potential
on the electron gun can be varied to change the energy of the emitted
electrons and there is a uniform axial magnetic field that guides the
electrons along the axis of the chamber. There is a toroidal magnetic coil (a
coil that is wound on a high permeability toroidal core such that the field
lines are confined within the torus and the magnetic lines do not appreciably
leak out) placed with its plane perpendicular the path of the electrons such
that all or most of the electrons pass through the toroid. Significantly,
there is no appreciable probability for the electrons to go `around' the coil,
and thus mostly they go through the toroidal hole. The authors reported that
the current detected at the plate is an oscillatory function of the magnetic
flux in the toroidal coil with the other parameters fixed. Since the magnetic
field cannot leak out into the path of the electrons, this was interpreted as
the effect of the vector potential (which is nonzero along the path of the
electrons, and which depends linearly on the current through the coil) on the
`phase' of the electrons in the same way one observes a phase shift in the
well known Aharonov-Bohm effect in quantum mechanics. The surprise here is the
fact that the electrons are not coherent over the scales of the experiment and
that the scale of the oscillations themselves correspond to equivalent
wavelengths of several centimeters! The de Broglie wavelength of the electrons
(with energy several 100 eV) is less than a nanometer, and the experiment is
done over a length scale of about 30 cm. Also, there aren't two interfering
paths, one through the toroid and one outside, to resemble the Aharonov-Bohm
geometry. Yet, the authors interpret the result as due to a Aharonov-Bohm
effect in the macrodomain, with the Larmour action playing the role of the
Planck's constant. They also assert that the effects indicate a violation of
classical electrodynamics. If true, then such a claim would imply a large
change in two of the most successful theories in physics. Therefore it is
important to critically examine the results of this experiment.

The interpretation of one dimensional interference in this type of experiments
has been discredited when such effects were reported earlier \cite{varmaold}.
The oscillatory patterns of currents in such macroscopic classical experiments
are entirely due to the focussing of electron beam in the axial magnetic
field. This was shown both experimentally and theoretically in references
\cite{unni1,unni2}. Multiple focussing of the electron beam from the source to
the detector creates a charge density pattern with an instantaneous spatial
distribution resembling a standing wave with the focus-to focus distance
determined by the axial velocity and the magnetic field. In fact, it is easy
to identify that the formula given by Varma \textit{et al} for the
``wavelength'', $\lambda=2\pi v_{||}/(eB/mc),$ is same as the standard
classical formula for the focussing distance of an electron beam with a small
angular spread propagating in a uniform axial magnetic field, $l_{f}=2\pi
v_{||}/(eB/mc)$. Once this is identified, all `macroscopic quantum effects'
reduce to classical effects arising from the focussed beam reaching the
detector from the source through a series of apertures and grids, affected by
the values of the magnetic field over the entire trajectory of the electrons
\cite{unni1,unni2}. Our analysis of the experimental results and of the
interpretation by Varma et al shows that their interpretation is based a lack
of appreciation of physical facts regarding interference phenomena, and also
classical motion of electrons in magnetic fields.

One important observation that reveals that the oscillatory patterns are due
to some effect due to the focussing of the electron beam in the magnetic field
is the estimate of the focussing length of the beam in the experiment. For a
monoenergetic electron beam with energy $E=\frac{1}{2}mv^{2},$ and small
angular spread, in a magnetic field $B,$ focussing occurs at the `focal
length',%
\begin{equation}
l_{f}=2\pi v_{||}/(eB/mc)
\end{equation}
where $v_{||}$ is the axial component of the velocity. Since the time period
for Larmour cycle is $T=2\pi/(eB/mc),$ the focal length is just the distance
travelled by the electrons during this time period. For a beam energy of 1200
eV, and an axial magnetic field of 2.70 mT used in the experiment, the focal
length is about 27 cm. \emph{Note that this is almost exactly the distance
between the source and the grid in front of the detector in the experiment}!
For focussing distance between 27--30 cm, the focus will fall close to either
the grid or the plate. Most interestingly, for the other energies and magnetic
fields for which the experiment was done, we get exactly the same focussing
distance within 2\% -- at 600 eV, and 1.89 mT , we get $l_{f}=27.3$ cm, at 800
eV and 2.25 mT, $l_{f}=26.5$ cm\ (Varma \textit{et al} wrongly estimate it to
be between 1 cm and 5 cm, and interpret this length as the `wavelength' of the
macroscopic matter waves, in their first footnote.) This shows unambiguously
that the results are closely related to the focussing of the electron beam on
the grid.

First, we show that the data presented by Varma \textit{et al} as evidence for
the classical A-B effect is inconsistent with the physics of the
interpretation of one dimensional interference. Interference or resonance
effects like that in Fabry-Perot cavity - in classical wave physics or in
quantum mechanics - happens due to the existence of multiple amplitude
interfering at the detection point. According to Varma \textit{et al} the two
interfering amplitudes correspond to the path between the electron source and
the detector plate (about 30 cm), and the path between the grid near the plate
and the plate itself (of the order of a cm). According to them the grid is a
source of forward scattered waves which interfere with the primary waves. The
wavelength of the macroscopic wavefunction, according to the authors, is
\begin{equation}
\lambda=2\pi v_{||}/\Omega
\end{equation}
where $\Omega=eB/mc.$ For a 1200 eV electron beam, this can be evaluated in
the axial field of about 2.7 mT used by Varma \textit{et al} to be 27 cm (We
have already shown that this is the focussing distance for the classical
electron beam, wrongly interpreted as a wavelength). This clearly shows that
there is no possibility of any appreciable amplitude of the wavefunction
between the grid and the plate since the wavelength of 27 cm is much larger
than the spatial separation which is only about 1 or 2 cm. This `wavelength'
is also much larger than the wire grid spacing, which makes the space between
the wire grid and the plate a forbidden region for the waves (see next
paragraph). Since the electron energy is much larger than the retarding
voltages on the grids, no primary electron can reflect back into the region of
the vector potential and execute multiple passages. Varma \textit{et al} write
the two amplitudes arbitrarily, going against the physical fact that there
cannot be appreciable amplitude between the grid and the plate when the
relevant wavelength is much larger (25-30 times in this case) than the region
of space in the problem.

Even if we assume that somehow one could formally write an amplitude for the
forward scattered wave after the grid, another severe problem arises, since
the forward scattering amplitude depends on the incident amplitude on the
grid, due to continuity. Unlike the usual configuration in an A-B effect
experiment, all the possible paths go through the toroid in this case. In a
genuine A-B effect experiment, one of the paths is within the toroid and the
other path is outside, bringing in the topology of a multiply connected region
into the problem. In the experiment by Varma \textit{et al}, this is not the
case, and the region in which the possible paths exist is simply connected.
This is contrasted in the schematic figure, Fig. 2.%

\begin{figure}
[ptb]
\begin{center}
\includegraphics[
height=1.7737in,
width=4.6631in
]%
{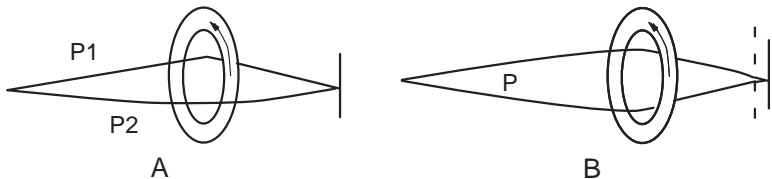}%
\caption{A) The configuration for an Aharonov-Bohm effect experiment. The
quantum amplitudes P1 and P2 encircle the toroid and they are in a multiply
connected region. B) The configuration in the experiment by Varma et al. All
paths P pass through the toroid, and the paths are in a simply connected
region.}%
\end{center}
\end{figure}

Another severe problem with their interpretation is related to the quantum
reflection properties of wave amplitudes with characteristic wavelength much
larger than the wire separation in the grid. Any consistent wave based
interpretation will show that waves with their wavelength much larger than the
grid wire spacing will either be absorbed or reflected back, and that the
transmission probability is very small. This means that for electrons with
macroscopic wavelength of 27 cm, a metal wire grid with spacings of the order
of millimeters is like an opaque metal plate! Therefore, Varma \textit{et
al}'s interpretation of macroscopic wavelike phenomena in this case is
severely flawed. The inconsistency is obvious.

There are other strong reasons to rule out the hypothesis of Varma \textit{et
al}. They admit the fact that the pattern is observed only when the axial
magnetic field is at a particular value for each electron energy, and that the
departure of even a few percent can wash out the pattern. If the effect was
genuinely due to the A-B like effect, this would not have been the case as we
will explain now. The condition for an `interference maximum', according to
Varma \textit{et al} is%
\begin{equation}
\Omega L=2\pi l\overline{v}%
\end{equation}
where $\overline{v}$ is the average axial velocity of the electron beam and
$l$ is an integer. $L$ is the distance between the source and the grid. If the
magnetic field deviates by a small amount from the condition for interference
maximum, then the interference pattern will be shifted by an amount
proportional to the change in the magnetic field due to the fact that Varma
\textit{et al} interprets the quantity $2\pi v_{||}/\Omega=2\pi mcv_{||}/eB$
as the wavelength $\lambda$ of the macroscopic wave. So, if the magnetic field
changes, the wavelength changes ($\delta\lambda/\lambda=-\delta B/B$) and the
interference pattern shifts by an amount equal to ($\delta\lambda/\lambda)L.$
Now, this shift can be compensated by a change in the `vector potential' if
there is a genuine A-B like effect. So, when the current is varied in the
solenoidal ring to change the vector potential, the oscillatory pattern should
show up, shifted by the appropriate amount, and with a slightly different
spacings between the peaks due to the slightly different `wavelength'. This is
the requirement for consistency. But Varma \textit{et al} fails to see any
oscillatory pattern when the magnetically field deviates as little as 5\% from
the value required to fulfill the equation above. Therefore there is no doubt
that the interpretation of macroscopic A-B effect is proved to be inconsistent
by their data itself.

Their observation that the effect is not seen when the magnetic field deviates
slightly from specific values give us a good clue as to the physical origin of
the oscillatory pattern seen by Varma \textit{et al}. As pointed out earlier,
the condition for `interference maximum' is nothing but the expression for the
focussing distance of the electron beam in the axial magnetic field. Therefore
the condition $\Omega L=2\pi l\overline{v}$ is same as the expression for the
`wavelength', $\lambda=2\pi v_{||}/\Omega,$ when $l=1.$ So, when the condition
$\Omega L=2\pi l\overline{v}$ is met for $l=1,$ the first focus point occurs
at the detector grid itself. (Varma \textit{et al} writes that the `typical
values for the wavelength' in their experiment are 1 -- 5 cm. But this is not
correct. As shown earlier, if the numerical values are substituted in the
expression for the wavelength, we get about 27 cm, which is exactly the
distance between the source and the grid in their experiment.) Now, it is easy
to see why a very small perturbation of the beam can create oscillatory
patterns at the detector. If the focus point is on the wire grid, part of the
beam will be easily blocked by the wire unless its diameter is much smaller
than the size of the focal point. Small perturbation can then shift the focal
point slightly, either axially or parallel to the grid, and this will cause
variations in the current. Such an obvious fact is not checked carefully in
their experiment.

There is also an error made by the authors regarding the estimate of the
Larmour radius of the beam. They state, in the caption to their figure 1, that
the beam diameter is about 2 mm, due to channelization by the magnetic field,
and therefore it is much smaller than the diameter of the toroidal solenoid
which is 2.6 cm in diameter. This is in error, and a proper estimate reveals
that the diameter of the beam is comparable to the diameter of the solenoid.
The Larmour radius for electrons in the magnetic field is given by
\begin{equation}
r_{L}=v_{\bot}/(eB/mc)
\end{equation}
where $v_{\bot}$ is the transverse component of the velocity of the electrons.
Since $v_{\bot}$ is approximately $v_{||}\sin\theta_{i}$ for a beam injected
at small angle $\theta_{i}$,
\begin{equation}
r_{L}\simeq v_{||}\sin\theta_{i}/(eB/mc)=l_{f}\sin\theta_{i}/2\pi
\end{equation}
The injection angle is upto 15$^{o}$ in their experiment and the maximum
Larmour radius of \ the beam can be estimated to be about 1.1 cm, for the
focal length of 27 cm we have already estimated. Thus, the diameter of the
beam is about 2.2 cm, about 10 times more than the estimate by Varma
\textit{et al}! Small misalignment amounting to a few degrees, or a few
millimeters, and distortion in the magnetic field due to the toroid core (as
evident in the magnetic field profile in their figure 1) can make the toroid
block a small part of the beam and affect the actual current reaching the
detector .

It is important to note that a physical blocking of the beam is not essential
for the perturbations of the detected current to occur. Since the pattern is
very sensitive to the applied field as Varma et al admit, even a tiny leakage
field can affect the electron trajectories. We note that there is significant
alteration of the axial magnetic field by the presence of the solenoidal ring
itself. The magnetic field along the axis is affected as much as 25\%,
presumably due to the magnetic field lines crowding through the high
permeability material. Whether the toroid is inside the chamber or outside,
the axial magnetic field lines are distorted considerably since they have to
pass through the high permeability toroid core, and appreciable perturbations
in the trajectory can occur. Instead of checking these obvious facts in the
experiment, the authors chose to jump to the conclusion that there was a new
discovery that would change two fundamental theories in a drastic way.

Also, whether the beam is focussed or not near an electrode affects the rate
of secondary electron emission from the electrode as discussed in reference
\cite{unni1}. These secondary electrons typically have low energy and get
accelerated back by the grid which is at a negative potential. When they
approach the high negative potential of the source electrode, another
reflection takes place and the secondary electrons reach back the grid and a
good fraction of them cross to the detector plate behind since their energy is
just sufficient to cross the grid (a few electron-volts above the grid
potential). These secondary electron can feel any leakage field from the
solenoidal coil twice in their passage back and forth, and are affected by
small stray fields. To test whether the contribution of secondary electrons to
the oscillatory pattern seen by Varma \textit{et al} is significant, \ more
diagnostic experiments are needed. Even if the contribution of the secondary
electrons is small, the modulations seen by Varma \textit{et al} could not be
due to an Aharonov-Bohm kind of effect, as we have argued on the basis of the
focussing of the primary beam itself. \ 

A recent report by the same authors on the observations of beat like phenomena
in the macroscopic domain and their interpretation in terms of macroscopic
quantum effects \cite{varmapre} have also been shown to be spurious and
completely explicable in terms of the standard classical scenario
\cite{unnipre}. It is to be stressed that all their previous observations of
wave-like effects, interpreted as due to macroscopic quantum phenomena by
Varma \textit{et al }\cite{var-physrep}, have been now shown to be due to
classical focussing and secondary electron generation. There is no compelling
reason to bring in any new physical effect to explain the observations. On the
contrary, explanation of the data in terms of wave effects leads to severe inconsistencies.

The reasoning given above is sufficient to rule out any claim that the
observed oscillatory patterns seen by Varma \textit{et al} is evidence for the
quantum like effect of a vector potential on the electron beam, analogous to
the A-B effect.

While many other shortcomings of the experiment can be pointed out, these are
not so important considering the fact that the interpretation in terms of
macroscopic quantum wave-like effects is shown to be entirely inconsistent. In
fact, it is surprising that Varma goes on to write that their results indicate
a subtle violation of the Lorentz force, whereas in reality the whole effect
could originate in Lorentz force on electrons in the magnetic field!

\end{document}